\documentclass[11pt,twoside]{article}  
\usepackage{asp2006}
\usepackage{adassconf}
\usepackage[dvips]{graphicx}
\begin{document}   

%

\paperID{E.16}

%

\title{ SuperAGILE data processing services }
       
%
%
%
%
%

\markboth{Francesco Lazzarotto et al}{SuperAGILE data processing services}

%




\author{Lazzarotto F., Costa E., Del Monte E., Donnarumma I., Evangelista Y., Feroci M., Lapshov I., Pacciani L., Soffitta P.}
\affil{National Institute for Astrophysics (INAF) IASF Rome, Italy}
\author{Trifoglio M., Bulgarelli A., Gianotti F.}
\affil{National Institute for Astrophysics (INAF) IASF Bologna, Italy}

\contact{Francesco Lazzarotto}
\email{francesco.lazzarotto@iasf-roma.inaf.it}

%
%

\paindex{Lazzarotto, F.}
\aindex{Costa, E.}
\aindex{Del Monte, E.}
\aindex{Donnarumma, I.}
\aindex{Evangelista, Y.}
\aindex{Feroci, M.}
\aindex{Lapshov, I.}
\aindex{Pacciani, L.}
\aindex{Soffitta, P.}
\aindex{Trifoglio, M.}
\aindex{Bulgarelli, A.}
\aindex{Gianotti, F.}


\keywords{space mission!ground segment!data analysis!pipeline}
\setcounter{footnote}{3}
\begin{abstract}          
The SuperAGILE (SA) instrument is a X-ray detector for Astrophysics measurements, part of the Italian AGILE satellite for X-Ray and Gamma-Ray Astronomy launched at 23/04/2007 from India. SuperAGILE is now studying the sky in the 18 - 60 KeV energy band. It is detecting  sources with advanced imaging and timing detection capabilities and good spectral detection capabilities. Several astrophysical sources has been detected and localized, including Crab, Vela and GX 301-2. The instrument has the skill to resolve correctly sources in a field of view of [-40, +40] degrees interval, with the angular resolution of 6 arcmin, and a spectral analysis with the resolution of 8 keV.
Transient events are regularly detected by SA with the aid of its temporal resolution (2 microseconds) and using signal coincidence on different portions of the instrument, with confirmation from other observatories. 
The SA data processing scientiﬁc software performing at the AGILE Ground Segment is divided in modules, grouped in a processing pipeline named SASOA. The processing steps can be summarized in \textit{data reduction}, \textit{photonlist building}, \textit{sources extraction} and \textit{sources analysis}. The software services allow orbital data processing (near real-time), daily data set integration, Temporal Data Set (TDS) processing and TDS processing with source target optimization (TDS\_SRC). Automatic data processing monitoring and interactive data analysis is possible from an internet connected workstation, with the use of SA data processing Web services. Many solutions were implemented in order to achieve fault tolerance. Archive management and data storage are performed with the help of relational database instruments.
\end{abstract}
\section{The SuperAGILE processing pipeline}
\begin{deluxetable}{|c|l|l|}
\tablecaption{SASOA software Data Processing Stages \label{E.16-tbl-1}}
\startdata
\textbf{\begin{tiny}n.\end{tiny}} &\textbf{\begin{tiny}STAGE\end{tiny}} & \textbf{\begin{tiny}Description\end{tiny}} \nl
\hline
\hline
\begin{tiny}1\end{tiny} & \begin{tiny}TRIGGER\end{tiny} & \begin{tiny}control if new data are transmitted\end{tiny}  \nl
\hline 
\begin{tiny}2\end{tiny} & \begin{tiny}DATA\_DOWNLOAD\end{tiny} & \begin{tiny}if triggered download available data\end{tiny} \nl
\hline
\begin{tiny}3\end{tiny} & \begin{tiny}COMMANDS\_PARSING\end{tiny} & \begin{tiny}parse run options\end{tiny} \nl
\hline
\begin{tiny}4\end{tiny} & \begin{tiny}LOCAL\_SETTINGS\end{tiny} & \begin{tiny}set hosts and data coordinates \end{tiny} \nl
\hline
\begin{tiny}5\end{tiny} & \begin{tiny}DATA\_FETCH\end{tiny} & \begin{tiny}load data input for actual run in local buffer \end{tiny} \nl
\hline
\begin{tiny}6\end{tiny} & \begin{tiny}FILE\_NAMING\end{tiny}& \begin{tiny}creates file names following the standard\end{tiny} \nl
\hline
\begin{tiny}7\end{tiny} & \begin{tiny}CORRECTION\end{tiny}& \begin{tiny}perform data correction task\end{tiny} \nl
\hline
\begin{tiny}8\end{tiny} & \begin{tiny}ORBIT\_REBUILDING\end{tiny}& \begin{tiny}rebuild data related on a given orbit\end{tiny} \nl
\hline
\begin{tiny}8\end{tiny} & \begin{tiny}DATA\_REDUCTION\end{tiny}& \begin{tiny}data separation, reconstruction and equalization\end{tiny} \nl
\hline
\begin{tiny}9\end{tiny} & \begin{tiny}PHOTONLIST\_BUILDING\end{tiny}& \begin{tiny}analyse data of the reduced photon list\end{tiny} \nl
\hline
\begin{tiny}10\end{tiny} & \begin{tiny}ATTITUDE\_CORRECTION\end{tiny}& \begin{tiny}correction for the attitude wobbling\end{tiny} \nl
\hline
\begin{tiny}11\end{tiny} & \begin{tiny}IMAGING\end{tiny}& \begin{tiny}image extraction \end{tiny} \nl
\hline
\begin{tiny}12\end{tiny} & \begin{tiny}FILING\end{tiny}& \begin{tiny}archive produced data\end{tiny}
\enddata

\end{deluxetable}
The SASOA (Super Agile Standard Orbital Analysis) pipeline is composed by a succession of elaboration tasks applied on the SA detected eventlist, the output of a certain task is the input of the following. Macro-tasks launch other subfunctions, some of them are implemented by external scripts and batch programs. In SASOA, a daemon running an endless loop provides to check if data for the incoming contacts are available and when they are found, the trigger for contact data processing is given. Data for a given contact mostly contain the data related to measurements acquired by the AGILE instruments during the last orbit. The failure of a single task of the pipeline may block the production of the final output data for a single run but the failure of a single contact run does not stop the daemon for the next incoming triggers and related data processing run. Different kinds of triggers can be used: the first way used to trigger the sasoa pipeline is to control if a file with extension .ok is generated by the preprocessing system for the incoming contact data, i. e. if a file named "VC1.002469.009.ok" is present, data for the contact 2469 are correctly generated and available on the data area of the preprocessing machine. Another trigger modality is to control if a value in a database of the ASDC Data Center is changed. When the system has assured that data are present, then the data are downloaded from the \textit{gtb} server in Bologna using the \textit{wget} utility (see M. Trifoglio et Al, “\textit{Archiving the AGILE Level-1 telemetry data}, Astronomy and Astrophysics, 2008”, ), data from ASDC are downloaded with \textit{wget}  also, but they have to be downloaded one by one (mirror and -A  option are disabled) to avoid transmission band overloading. Data products (histograms, plots, statistics \& reports) are generated at different levels, hosts coordinates, data directories paths and other configurations of the processing system are set. The most important input telemetry data used for SA instrument data analysis are eventlist data (SA TM 3905), attitude data (SA TM 3914), and ephemerides data (SA TM 3916).
The file names for i/o file of all tasks are generated following the official naming convention, the correction stage performs time synchronization among input data files and other corrections. Then some actions useful to rebuild the orbital organization of the data are performed: if the contact contains more than 1 orbit, then data are divided in the respective parts and the pipeline is launched on the first part of data. After the run end, the system is triggered on the 2nd part. In the Data Reduction stage the eventlist from the previous stage is taken as input (complete orbit or orbital part), with efficiency files and then a reduced output eventlist is generated. In the Sources Extraction stage, the data are corrected removing wobbling effects, then the exposure is evaluated using informations regarding Earth Occultation and SAGA time periods. After these steps the imaging procedure is applied and sources lists are generated, giving position and flux for the detected sources.  \\
\begin{figure}[t]
\epsscale{0.70}
\plotone{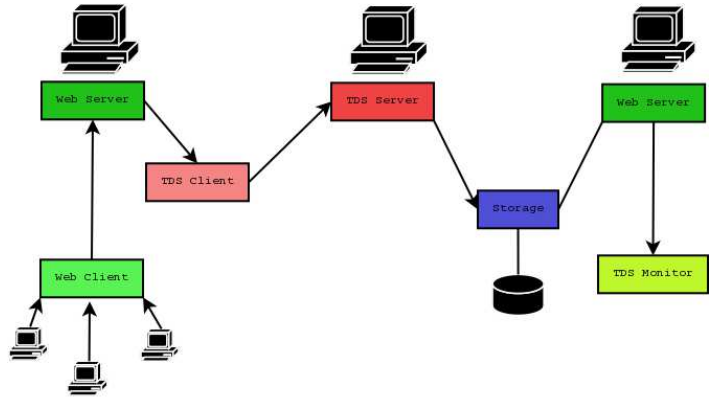}
\caption{TDS processing schema} \label{E.16-fig-1}
\end{figure}
\begin{figure}[t]
\begin{center}
\epsscale{0.70}
\includegraphics[angle=-90, width=8cm]{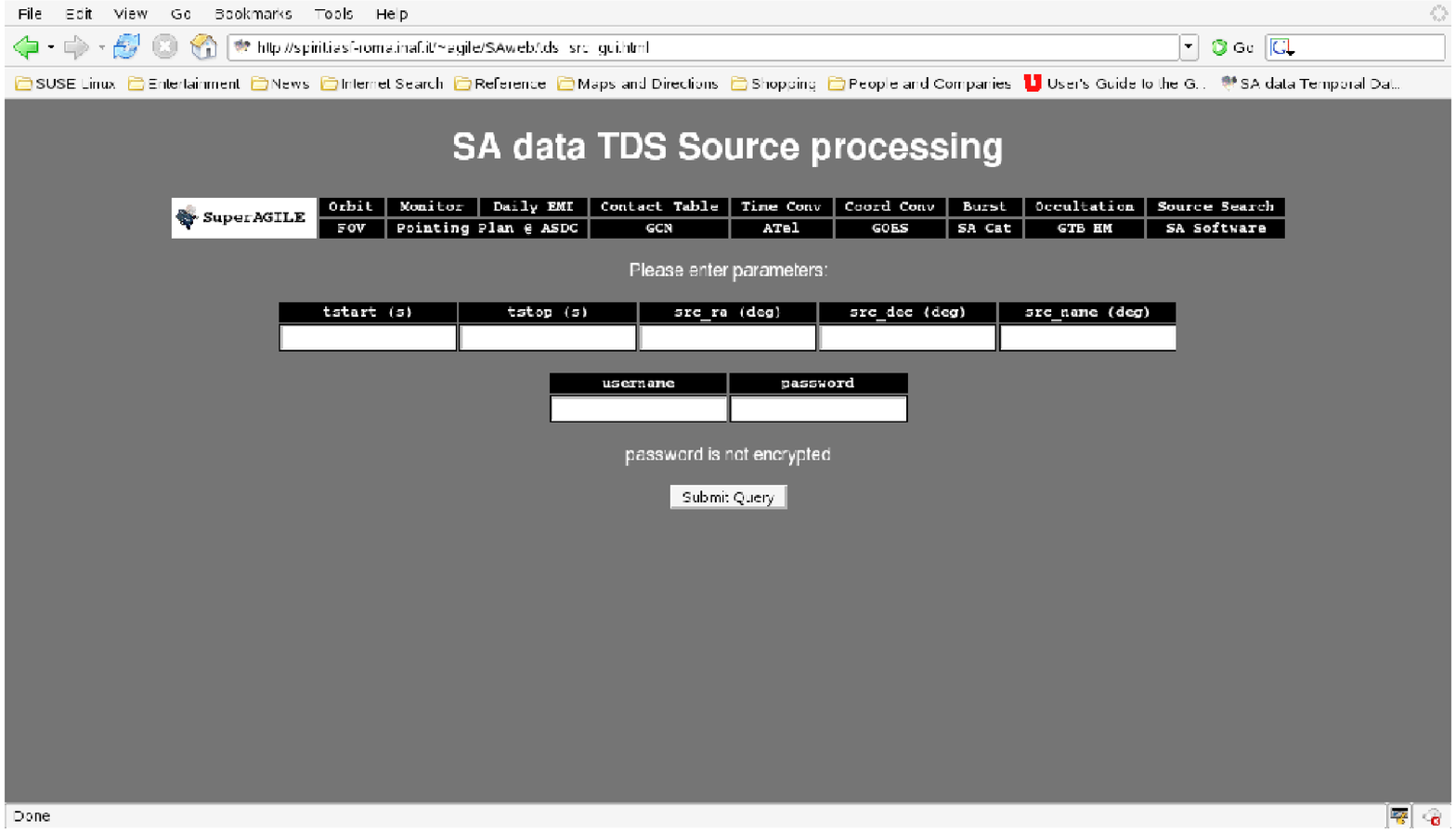}
\caption{TDS SRC software GUI} \label{E.16-fig-2}
\end{center}
\end{figure}
\begin{figure}[t]
\epsscale{0.70}
\plotone{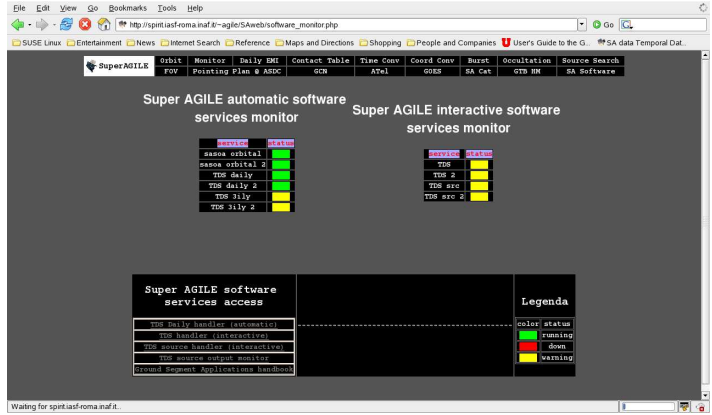}
\caption{SA software monitor} \label{E.16-fig-3}
\end{figure}
\section{Temporal Data Set processing}
Telemetry data are preprocessed and archived with the aid of the \textit{tmpps} system furnished by the GTB Team (see M. Trifoglio et Al, Astronomy and Astrophysics, 2008). Data of SuperAGILE measurements have to be integrated over days and weeks  to extract science information. The software utility performing this task is named TDS processing (TDS). The working schema of TDS is shown in fig. \ref{E.16-fig-1}.
A web graphical user interface (GUI) is provided to launch data analysis applications through the web, the scientific user can perform its own analysis from elsewhere and using the preferred operating system (fig. \ref{E.16-fig-2}).
TDS processing can be refined with specific source analysis operations, optimizing the results calculating the exposure of the celestial source in the given position, that is an input to the computation, expressed with (ra, dec) coordinates. This has also allowed long integration analysis on 3C 273 (see L. Pacciani et al, Astronomy and Astrophysics, 2008) and on Markarian 421 (see  I. Donnarumma et Al,
“\textit{The June 2008 flare of Markarian 421 from optical to TeV energies}”,  ApJ, 2008).
A web based monitor is showing the status of the automatic processing software in Figure~\ref{E.16-fig-3} .
\section{Conclusions}
The SA scientific processing system is ready, enabling the access from web solve installation \& portability problems.
Processing stations will be added to fit with the actual number of data analysis software users. 
All SA software is portable, we have in use running versions of the whole processing system under Linux OpenSUSE (until 10.3 version, for 32 \& 64bit architecture).


\end{document}